\documentclass[runningheads]{llncs}
\usepackage[T1]{fontenc}
\usepackage{amsmath}
\usepackage{algpseudocode}
\usepackage{algorithm}
\usepackage[labelformat=simple]{subcaption}

\usepackage{graphicx} 
\usepackage{soul}
\usepackage{color}
\usepackage{amssymb}
\usepackage{authblk}
\usepackage{comment}
\usepackage{xcolor}
\usepackage[inkscapelatex=false]{svg}

\allowdisplaybreaks

\DeclareMathOperator*{\argminA}{arg\,min}

\usepackage[compress]{cite}

\usepackage{graphicx}

\begin{document}
%
\title{Distributed Convolutional Neural Network Training on Mobile and Edge Clusters}

\titlerunning{Distributed CNN Training on Mobile and Edge Clusters}

\author{Pranav Rama \and
Madison Threadgill \and
Andreas Gerstlauer}
\authorrunning{Pranav Rama et al.}

%
\institute{Electrical and Computer Engineering\\
The University of Texas at Austin, Austin TX, USA\\
\email{\{pranavrama9999, madison.threadgill, gerstl\}@utexas.edu}}
\maketitle              
\begin{abstract}
The training of deep and/or convolutional neural networks (DNNs/CNNs) is traditionally done on servers with powerful CPUs and GPUs. Recent efforts have emerged to localize machine learning tasks fully on the edge. This brings advantages in reduced latency and increased privacy, but necessitates working with resource-constrained devices. Approaches for inference and training in mobile and edge devices based on pruning, quantization or incremental and transfer learning require trading off accuracy. Several works have explored distributing inference operations on mobile and edge clusters instead. However, there is limited literature on distributed training on the edge. Existing approaches all require a central, potentially powerful edge or cloud server for coordination or offloading. 
In this paper, we describe an approach for distributed CNN training exclusively on mobile and edge devices. Our approach is beneficial for the initial CNN layers that are feature map dominated. It is based on partitioning forward inference and back-propagation operations among devices through tiling and fusing to maximize locality and expose communication and memory-aware parallelism. We also introduce the concept of layer grouping to further fine-tune performance based on computation and communication trade-off.
Results show that for a cluster of 2-6 quad-core Raspberry Pi3 devices, training of an object-detection CNN provides a 2x-15x speedup with respect to a single core and up to 8x reduction in memory usage per device, all without sacrificing accuracy. Grouping offers up to 1.5x speedup depending on the reference profile and batch size. 

\end{abstract}

\section{Introduction}
Traditionally, training and inference of deep learning (DL) models is performed in the cloud. This requires a large amount of data to be collected and sent to a centralized infrastructure, introducing latency, privacy, and real-time concerns. Various approaches have proposed to partition the processing between mobile, edge, and cloud resources~\cite{ren_survey_2022, Yousefpour_2019}.  However, such approaches still rely on a remote cloud for partial processing. To address the latency and privacy concerns when communicating with the cloud, recent efforts have emerged to localize DL tasks fully on mobile or edge devices \cite{stahl_deeperthings_2021, zhou_adaptive_2019, du_distributed_2020}. However, this brings the challenge of performing compute and memory-intensive inference and training operations on such resource-constrained devices.


A wide range of approaches have been proposed to address limited memory and computing capabilities in mobile and edge settings. Techniques such as pruning and quantization focus on decreasing the complexity of the model by removing weights and neurons or reducing the bit precision during inference and/or training.
Other approaches such as incremental and transfer learning start from a pre-trained model and only partially update the model to save computational resources and reduce training time.
These approaches trade-off accuracy for decreased computational complexity. 

Several complementary methods have recently been proposed to utilize parallelism in DL models by partitioning them across multiple devices while preserving the original model and accuracy. Federated learning \cite{li_federated_2020} exploits data parallelism, but still requires a central server for coordination as well as storing and processing of complete models in each device, which is often infeasible given memory constraints. Other approaches partition and distribute the model itself across a cluster of edge devices \cite{zhao_deepthings_2018, mao_modnn_2017, du_distributed_2020, zhang_deepslicing_2021}. In addition to exploiting available multi-device parallelism, this allows for reducing both the computational and storage requirements on each device. However, such approaches have only been demonstrated for inference so far. 

In this paper, we present an approach for distributed CNN training exclusively on communication- and memory-constrained mobile and edge clusters. Our approach targets feature map-dominated early CNN layers. We adopt a tiling and fusing-based partitioning scheme that has previously been demonstrated for inference \cite{fused_sb, zhao_deepthings_2018, stahl2023fused} and extend it to apply to both forward and back-propagation training tasks. The scheme tiles feature maps to reduce memory footprint and expose model parallelism, then fuses matching tiles of consecutive layers into independent execution stacks placed on each device to maximally exploit locality. Furthermore, groups of layers are formed among tiles where synchronization of feature data shared among neighboring tiles is performed only at group boundaries. At the end of a single training pass, the final weight updates of all stacks are aggregated. This approach can confirm to arbitrary memory constraints imposed by each edge device while exposing parallelism, minimizing communication, and exploiting the locality inherent in convolutional and pooling layers. Our distributed training approach includes the following contributions:

\begin{enumerate}
  \item We propose a novel method for tiling and fusing of backpropagation tasks that considers memory and communication constraints, while exploiting parallelism for distributed CNN training on resource-constrained device clusters.
  
  \item We apply the concept of layer grouping
  of forward inference and back-propagation tasks in order to further fine tune computation and communication overhead based on the grouping profile of the layers.

  \item We evaluate our approach on distributed training of Yolov2, a common CNN for object detection, distributed across a network of quad-core Rasberry Pis. 
  
\end{enumerate}


\section{Related Work}

Performing inference on resource-constrained edge and mobile devices has received significant attention. 
Approaches for distributed inference in edge settings 
exploit inherent parallelism to partition a model and distribute it across multiple devices~\cite{mao_modnn_2017, zhang_deepslicing_2021, du_distributed_2020, zhao_deepthings_2018, zhou_adaptive_2019}. These methods can be applied to the forward inference pass in distributed training. We adopt tiling and fusing strategies from distributed edge and hardware accelerated inference \cite{fused_sb, zhao_deepthings_2018, stahl2023fused} in our work and extend it to the back-propagation pass in order to support distributed training.   

Training CNNs requires additional memory compared to inference due to the need to store input data, gradients, and activation values for each layer.
This normally requires partitioning of the workload involving powerful edge servers or the cloud \cite{ren_survey_2022, noauthor_coopfl_nodate,sen2022distributed}.
Multiple approaches exist for training a DL model on a single edge device \cite{tinyml_review}. These typically employ simplified model architectures \cite{small_model} or use reduced bitwidths for training \cite{lin_-device_2022}. Alternatively, approaches for incremental or transfer learning take a pre-trained model and only update a subset of weights \cite{cai_tinytl_2021} or the last layers of a model with every training sample \cite{chiang_mobiletl_2023}. However, all of these approaches trade off accuracy for reduced model complexity.

Multi-device solutions that rely on federated learning exploit data parallelism to collaboratively train a model, with each device training a local model on its own data and devices exchanging weight updates as different variants of a distributed gradient descent \cite{li_federated_2020}. Other multi-device approaches rely on approximate gradient prediction methods that trade off accuracy \cite{chen_exploring_nodate}. However, mobile and edge devices, e.g. in the IoT space, often lack sufficient memory to store an entire local model. In \cite{noauthor_coopfl_nodate, sen2022distributed}, federated learning is hierarchically combined with model partitioning in each local cluster. However, these approaches use
partitioning schemes commonly used in cloud settings, which are not 
optimized for the greater memory and communication limitations in mobile and edge settings. 

\section{Overview}

\begin{figure}[tp]
\centering
  \hfill
  \begin{subfigure}{0.35\textwidth}
    \centering
    \includegraphics[width=.83\linewidth]{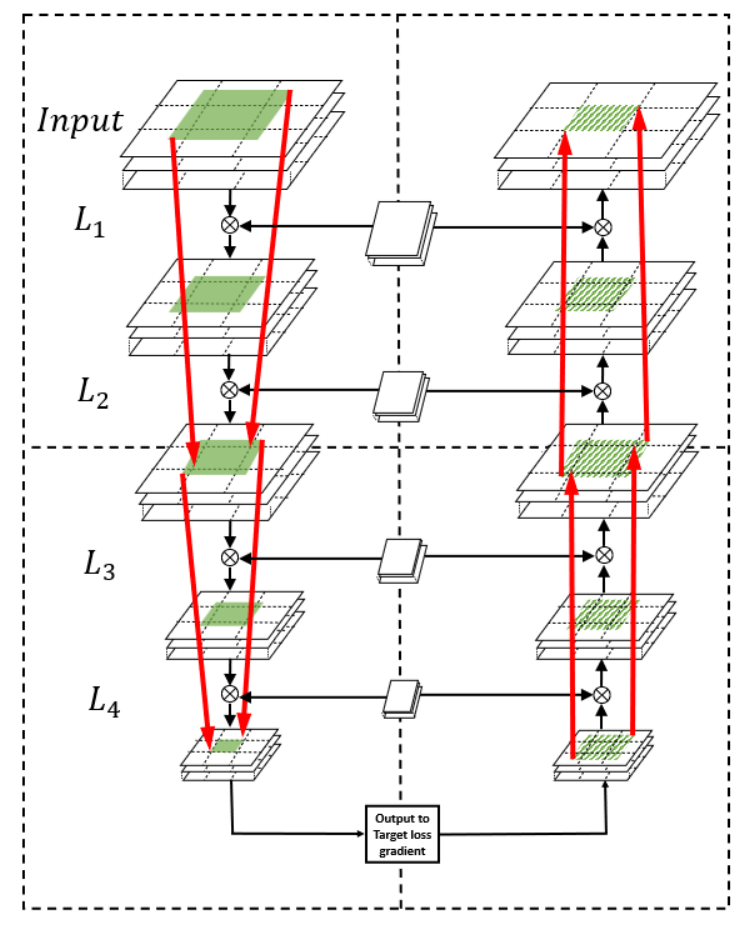}
    \caption{3x3 tiling w/ 2 groups} \label{fig:complete_net}
  \end{subfigure}%
  \hfill  
  \begin{subfigure}{0.5\textwidth}
    \centering
    \includegraphics[width=\linewidth]{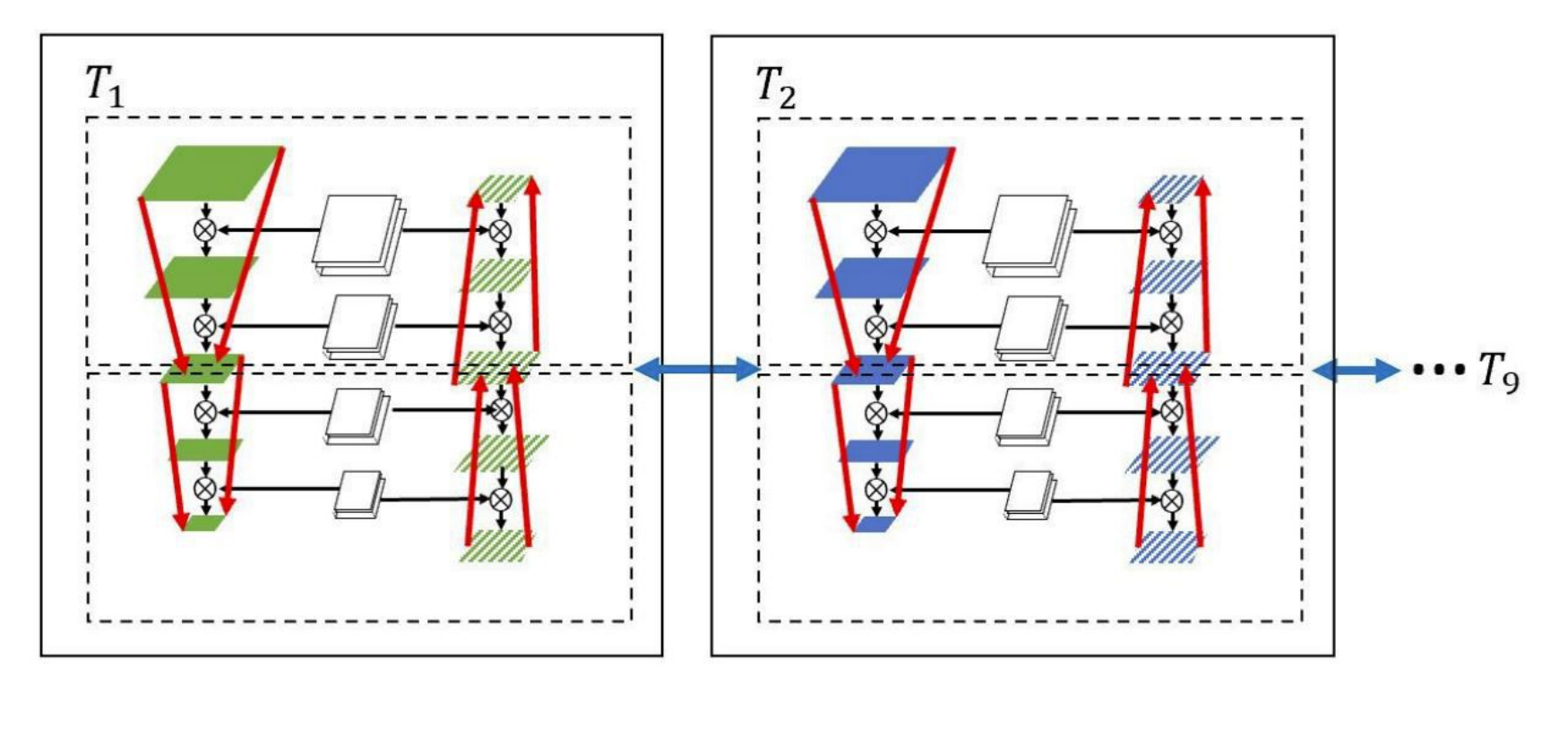}
  \caption{Independent execution tasks with communication boundaries} 
  \label{fig:partitioned_net}
  \end{subfigure}%
  \hfill
\caption{Distributed CNN training overview.} \label{fig:grouping}
\vspace{-10pt}
\end{figure}

Fig.~\ref{fig:grouping} gives an overview of our approach. 
We partition feature data and delta gradient maps in forward inference and back-propagation passes, respectively, into tiles in a grid-wise fashion along their width and height dimensions. In both passes, output tiles of each layer are computed from input tiles through convolutions with filter data or through simple pooling operations. Exploiting the inherent locality in these operators, all intermediate matching tiles on forward and backward passes are fused into independent execution stacks and tasks that stay local on one device. Tiling exposes parallelism and reduces storage requirements proportional to the tiling granularity, while fusing maximizes locality and thus minimizes communication overhead. 

Each output tile is computed by convolving a certain dependent input region with the filter data. This dependent region includes the corresponding input tile along with some boundary data, which depends on the filter size and stride. The boundary data has to be communicated between the neighboring devices prior to starting the convolutions in both the forward and backward passes.

Alternatively, we can further combine multiple convolutional and pooling layers to form groups where communication of boundary data with neighboring tiles is done only at the beginning of each group. Within each group, any required intermediate data is locally computed from input data collected at the beginning of the group and no further communication is needed within the group. Fig.~\ref{fig:grouping} shows 2 groups each in the forward and backward pass. In this case, communication is done at the feature-map inputs of layer $L_{1}$ and $L_{3}$ in the forward pass and at the delta gradient inputs of layer $L_{4}$ and $L_{2}$ in the backward pass. These feature and gradient maps serve as synchronization points where all tiles share boundary data with the neighboring tiles. Grouping introduces a trade off between communication and computation overhead. Larger groups have more redundant computation since the boundary data grows with the group size as illustrated by the funneling red arrows in Fig.~\ref{fig:grouping}. At the same time, larger groups synchronize less frequently whereas smaller groups have more communication and synchronization overhead. We will discuss optimal grouping strategies later. 

The only points at which the entire partition needs to be communicated is when receiving the input training sample at the first layer and the initial delta gradient loss at the last layer. Once this is received, the forward pass and backward pass can be completed with just intermediate group boundary synchronization, which is a much smaller overhead.

Each task and device requires access to a complete copy of all filters. In order to update the filter weights during back-propagation, partial weight gradients computed by each task for each tile must be summed across all tiles to get the final weight updates. This requires the devices to communicate their entire partial weight update sets with each other or a common central device for summing at the end of the training cycle for each batch. Such weight updates are only required once at the end of each batch, and can stay local on each tile until then. For the early feature-map dominated layers, filters are relatively small and storing local copies in each device as well as communicating updates between tiles carries a small overhead in comparison to the computation and memory benefits we get from feature and gradient map partitioning. 

\section{Distributed Training}
In the following, we describe details of our distributed tiling approach for a single layer followed by a discussion of fusing and grouping across multiple layers.

\subsection{Single-Layer Tiling} \label{sssec:slt}

\begin{figure}[tp]
    \centering
    \includegraphics[width=0.75\textwidth]{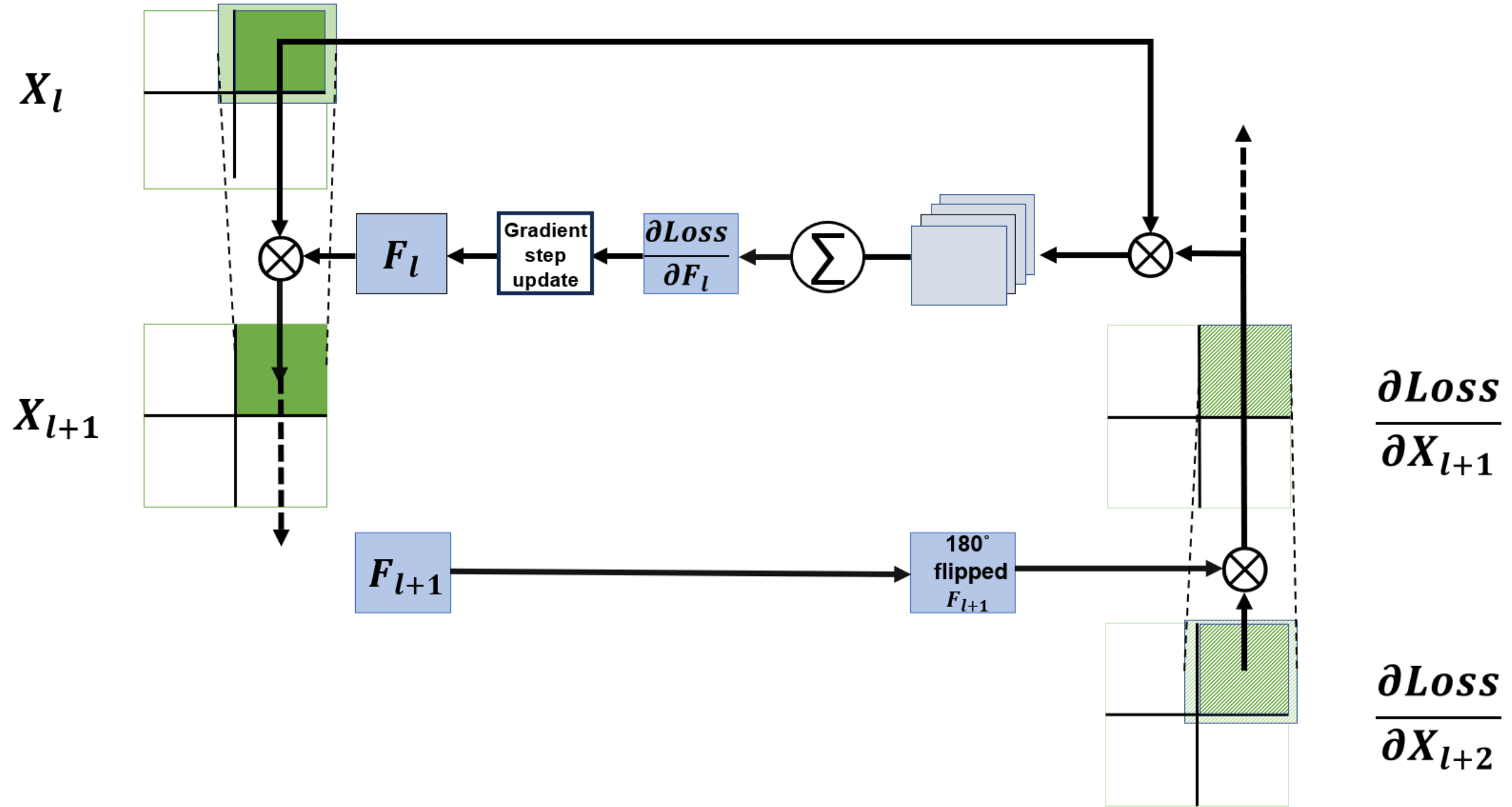}
    \vspace{-5pt}
    \caption{Single-layer tiled forward inference and back-propagation.}
    \label{fig:single_layer_details}
\end{figure}

Fig.~\ref{fig:single_layer_details} illustrates the tiling process of a forward pass, backward pass and weight update at layer \textit{l} for a 2x2 tile partition. $X_l$ and $X_{l+1}$ are the input and output feature maps of layer \textit{l}, respectively. Assuming a tiling into an NxM grid, in the forward pass, each of the tiles in $X_l$ with the necessary boundary data are convolved with the filter $F_l$ to produce the NxM tiled output feature-maps $X_{l+1}$. 

For back-propagation, we need to compute two gradients, the delta loss gradients and the weight updates. The delta loss gradients are obtained through recursive back-propagation starting with the loss gradients at the output of the last layer. To calculate the loss gradients $\frac{\partial Loss}{X_{l+1}}$, each output tile of the next layer's loss gradients, $\frac{\partial Loss}{X_{l+2}}$, together with the necessary boundary data is convolved with the $180^\circ$ rotated filter to produce the corresponding tile in $\frac{\partial Loss}{X_{l+1}}$. 

Finally, to compute the weight updates, the feature map tiles of $X_{l}$ are first convolved with the corresponding tiles of the delta loss gradient $\frac{\partial Loss}{X_{l+1}}$ to produce NxM filter gradient sets, one for each tile. These NxM weight updates are partial sums pertaining to the region of the map the tile is associated with, and the final weight gradients $\frac{\partial Loss}{F_{l}}$ can simply be obtained by summing them up.

This final gradient can then be used to update $F_{l}$ as illustrated in the figure. The summation requires each device to communicate their partial sums to a common device that performs the summation and transmits the updated weight gradients back to each tile. To minimize overhead, the summation can be done once for all filters in all layers at the end of the training cycle of a single batch.

\subsection{Fusing and Grouping} \label{sssec:fandg}

\begin{figure}[tp]
\hfill
  \begin{subfigure}{0.45\textwidth}
    \centering
    \includegraphics[width=0.5\textwidth]{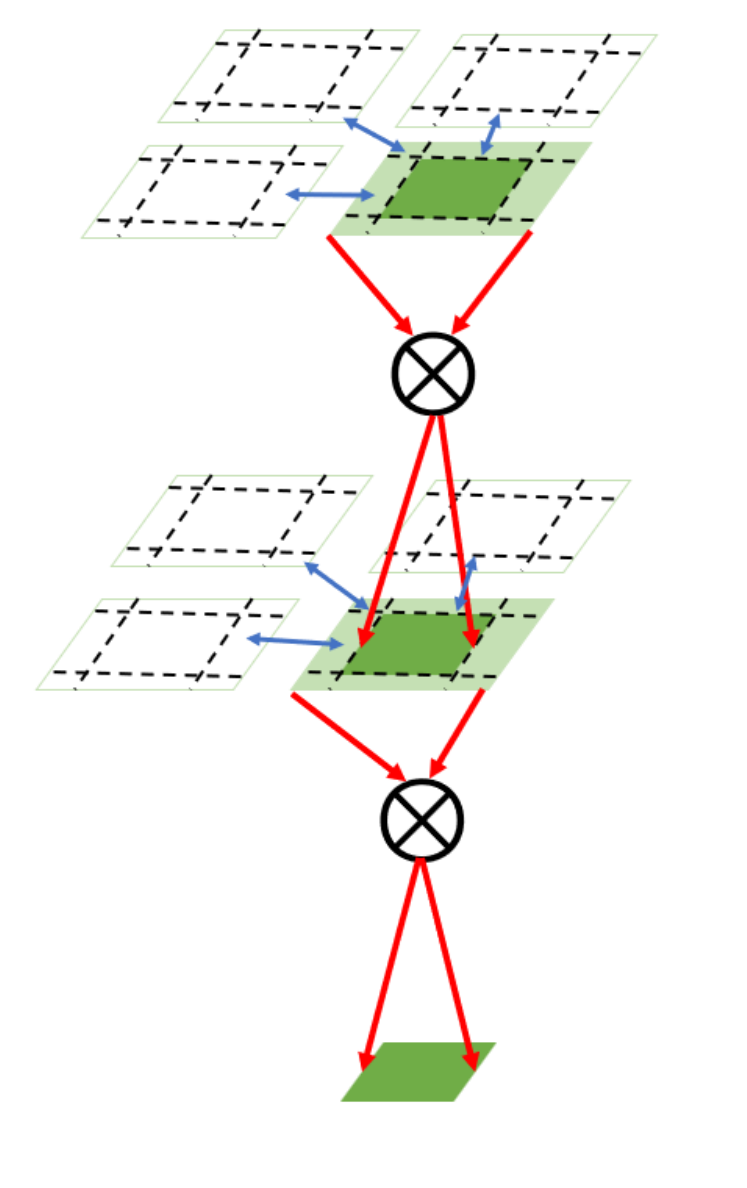}
    \vspace{-5pt}
    \caption{Fusing without grouping} \label{fig:fusing_illustration_2g}
  \end{subfigure}%
\hfill
  \begin{subfigure}{0.45\textwidth}
    \centering
    \includegraphics[width=0.5\textwidth]{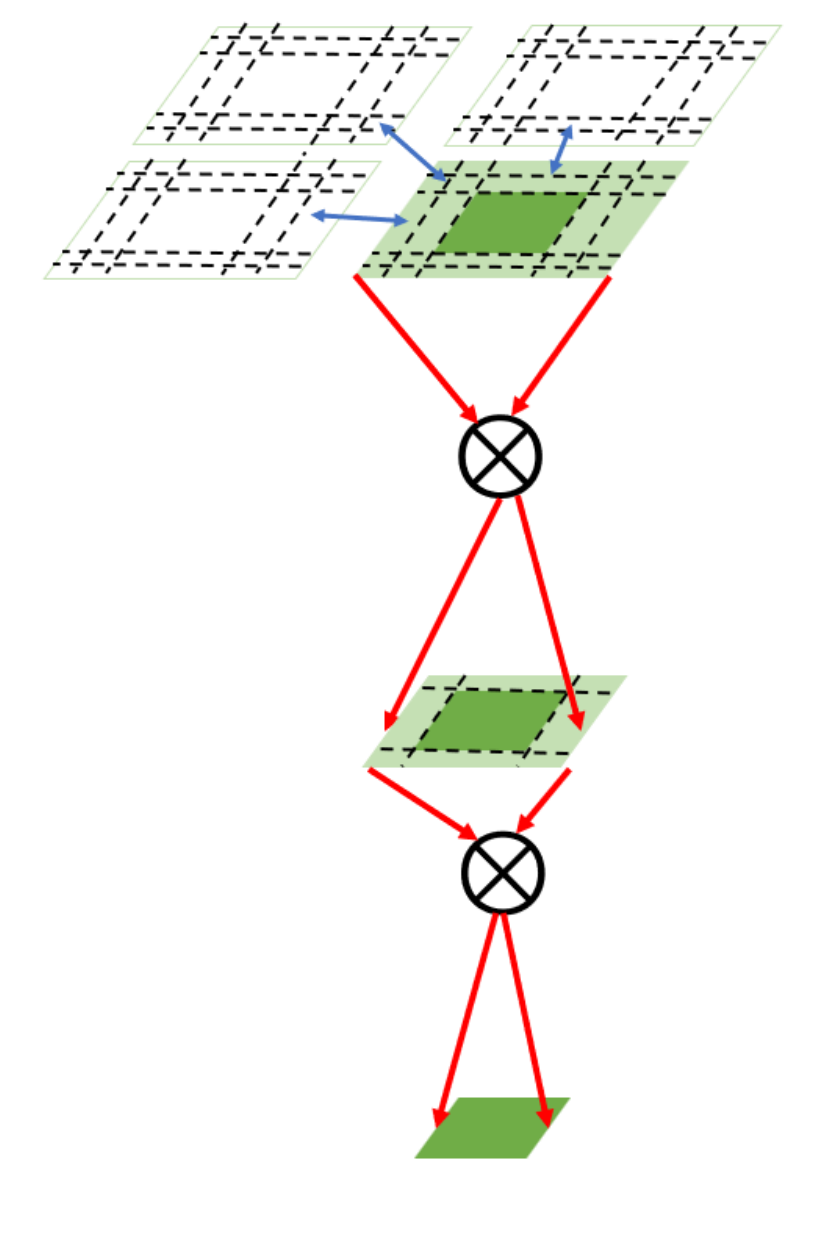}
    \vspace{-5pt}
    \caption{Fusing with grouping} \label{fig:fusing_illustration_1g}
  \end{subfigure}%
\hfill
\caption{Fusing and grouping illustration.} \label{fig:fusing_illustration}
\end{figure}

As introduced earlier, matching tile partitions in the forward and backward passes are fused across all convolutional and pooling layers in that they stay local on the same device. Fig.~\ref{fig:fusing_illustration} illustrates the fusing and grouping across 2 layers for a forward pass (the backward pass is symmetrical). The center region of each tile (dark green) is fused across layers, stays local on the device and is never exchanged with other devices (in both forward and backward passes).  However, the devices also exchange some neighboring boundary data (light green portion) required to complete the convolutions/pooling. Fig.~\ref{fig:fusing_illustration_2g} shows the case without grouping where boundary exchange occurs at the input to both layers thus having minimal redundant computation and storage. By contrast, Fig.~\ref{fig:fusing_illustration_1g} shows the case where the exchange only occurs at the input to first layer. However, in this case, the amount of shared boundary data per tile increases leading to more storage and redundant computation on each tile. 
In other words, grouping reduces the redundant computation and storage at the expense of additional communication and synchronization overhead, i.e. there is a trade-off between computation and communication.

\begin{figure}[tp]
    \centering
    \includegraphics[scale=.18]{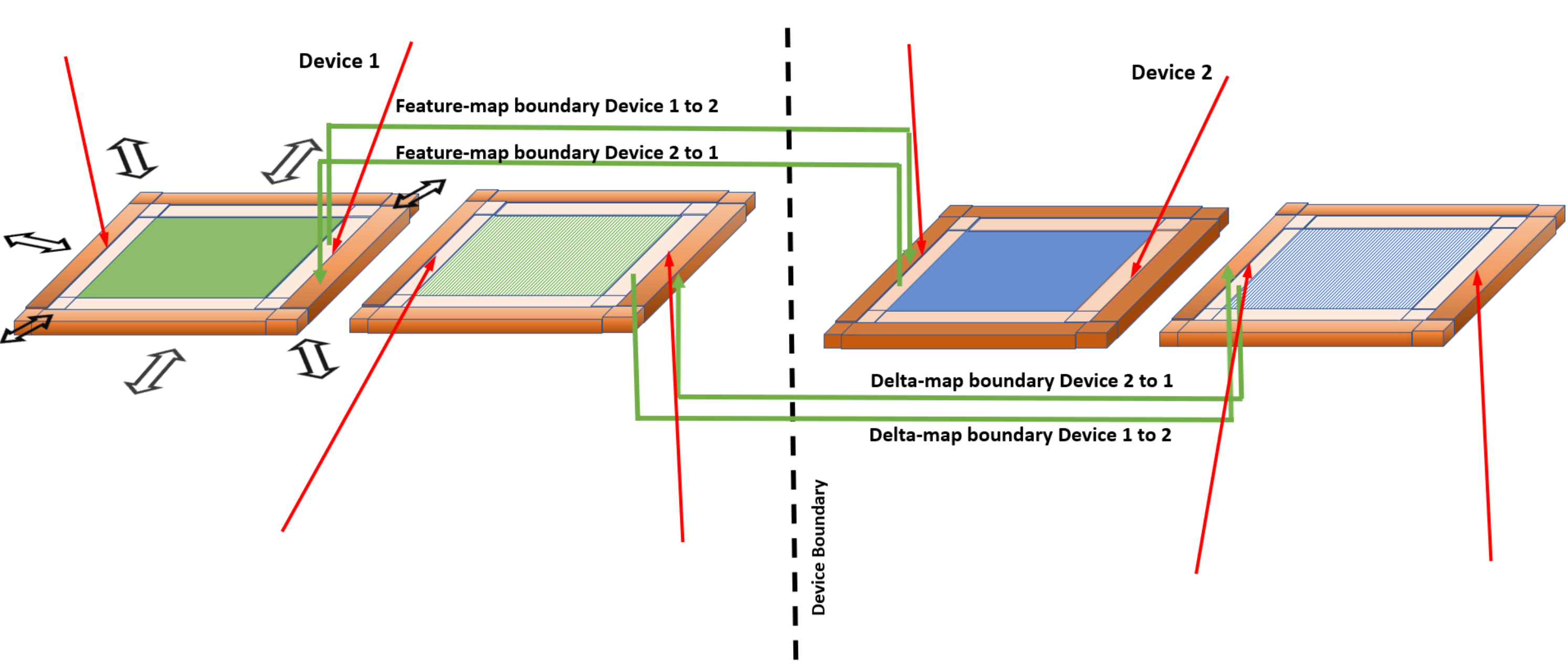}
    \vspace{-5pt}
    \caption{Group boundary communication illustration.} 
    \label{fig:group_boundary}
\end{figure}

Fig.~\ref{fig:group_boundary} illustrates a granular view of communication at the group boundary. Each device both transmits and receives the required boundary data to/from up to 8 neighboring tiles, where devices transmit data from the internal border of the locally computed tile while receiving boundary data external to it. These exchanges happen at the group input layers in forward and backward passes. 
The double ended arrows at the feature map in device 1 indicate that similar exchanges occur with the other 7 neighboring tiles, if present. The same happens at group input delta maps and feature maps across all tiles.

The span of each tile $(i,j)$ with boundary data at layer $l$ can be represented by its top-left $(x1_{l, (i, j)}, y1_{l, (i, j)})$ and bottom-right $(x2_{l, (i, j)}, y2_{l, (i, j)})$ co-ordinates. 
Furthermore, we represent the filter/kernel size and stride at layer $l$ as $K_{l} \times K_{l}$ and $S_{l}$, respectively. A group starting at the input to layer $s$ and ending at the input to layer $e$ (output of layer $e-1$) is represented as a tuple $(s, e)$.
 

Suppose we have a grid of tiles and want to create the grouping profile in the forward pass. To derive required boundary and tile data, we begin at the feature-map output of the last layer of the last group and recursively traverse backward among intermediate layers and groups. 
For any group, $(s, e)$, the feature map output of the last layer of the group, $e$, is partitioned length and breadth wise equally among the tiles. Then, we recursively compute the dependent region in the previous layers to produce the required feature map for each tile in each intermediate layer $l$ within the group, where $s < l \le e$. Given the tile co-ordinates at the input to layer $l$ (output of layer $l-1$), the required tile region at the input to layer $l-1$ is
    \begin{subequations}
      \begin{gather}
        x1_{l-1, (i, j)} = x1_{l, (i, j)} \times S_{l-1} - \lfloor \frac{K_{l-1}}{2} \rfloor \\
        y1_{l-1, (i, j)} = y1_{l, (i, j)} \times S_{l-1} - \lfloor \frac{K_{l-1}}{2} \rfloor \\   
        x2_{l-1, (i, j)} = x2_{l, (i, j)} \times S_{l-1} + \lfloor \frac{K_{l-1}}{2} \rfloor + (S_{l-1}-1) \\
        y2_{l-1, (i, j)} = y2_{l, (i, j)} \times S_{l-1} + \lfloor \frac{K_{l-1}}{2} \rfloor + (S_{l-1}-1)
      \end{gather}
    \end{subequations}   
    for convolutional layer $l-1$.

For the backward pass, computing group boundary data bounds is similar except that we go in the opposite direction, i.e. we start computing the co-ordinates from the delta gradient map output of the first layer of the network. For any group, $(s, e)$, given the tile co-ordinates of the delta map at intermediate layer $l$, where $s \le l < e$, the tile co-ordinates of the delta map at layer $l+1$ are
    \begin{subequations}
      \begin{gather}
        x1_{l+1, (i, j)} = \lceil \frac{x1_{l, (i, j)} - \lfloor \frac{K_{l}}{2} \rfloor}{S_{l}}  \rceil \\
        y1_{l+1, (i, j)} = \lceil \frac{y1_{l, (i, j)} - \lfloor \frac{K_{l}}{2} \rfloor}{S_{l}}  \rceil \\
        x2_{l+1, (i, j)} = \lfloor \frac{x2_{l, (i, j)} + \lfloor \frac{K_{l}}{2} \rfloor}{S_{l}}  \rfloor \\
        y2_{l+1, (i, j)} = \lfloor \frac{y2_{l, (i, j)} + \lfloor \frac{K_{l}}{2} \rfloor}{S_{l}}  \rfloor
      \end{gather}
    \end{subequations}
    for convolutional layer $l$.



After completing the forward and backward passes, the partial filter gradients are computed by convolving the corresponding delta gradients with the feature-maps as described in Section ~\ref{sssec:slt}. For this, just $\lceil \frac{K_{l}}{2} \rceil$ element wide boundary data would be required in the feature-map at layer \textit{l}. However, this data is already gathered during the forward pass and can be re-used to avoid additional communication for this step.

\section{Experiments and Results}
We implemented our distributed training approach in C on top of the Darknet framework 
and validated our model using the first 16 layers of the Yolov2 CNN. A reference implementation is available at~\cite{repo}. Our primary experimental test-bed consisted of 6 Raspberry-Pi3 devices with quad-core ARM Cortex-A53 CPUs and 1 GB of RAM each running a Linux kernel. Each tile was executed as an individual Linux process and we allocated upto 4 tiles per device to run on the 4 cores. The devices were all part of a local 100Mbps Ethernet network. For communication between processes within the same device, we used shared memory and local sockets to minimize overhead. TCP network sockets were used to communicate between processes across devices on the network. More details and results can be found in~\cite{techreport}.

\begin{figure}[tp]
    \centering
    \includegraphics[width=0.85\textwidth]{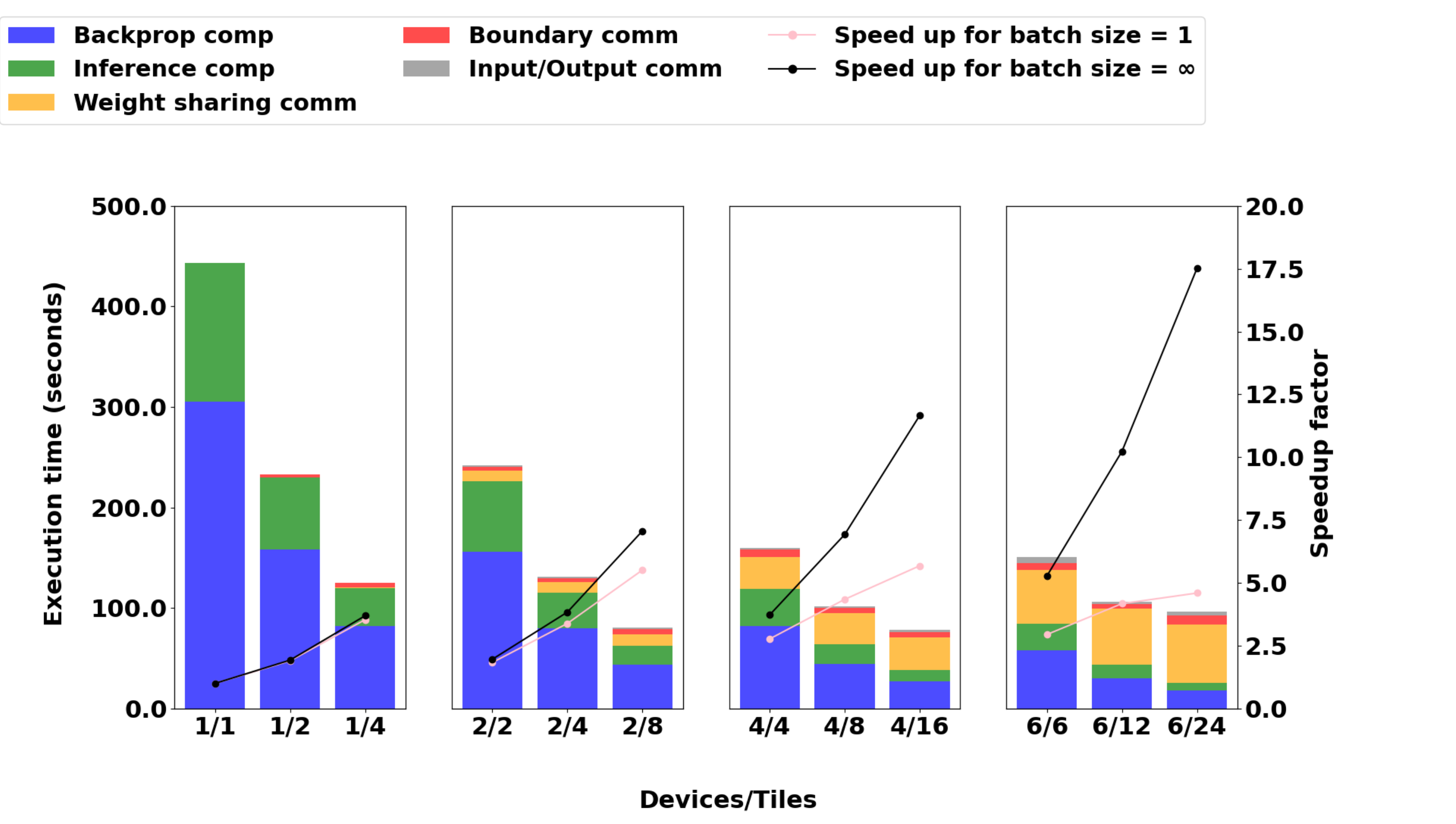}
    \vspace{-5pt}
    \caption{Execution time split and speedup with number of tiles and devices.}
    \label{fig:speedup}
\end{figure}

\subsection{Speedup}
Fig.~\ref{fig:speedup} shows the execution times for a single training sample (batch size of 1) across different combinations of devices and cores ranging from 1 to 6 devices, each using  1 to 4 of their cores. The number of tiles in a given device/core combination is the total number of cores across all devices. Each tile was scheduled as an independent process. Results are broken down into execution times for back-propagation and forward inference computations, communication times for filter weight updates and boundary data exchanges, and input/output communication overheads.
A single device with 1 tile (1 process - single core) took around 7 minutes to finish the training cycle (forward pass, backward pass and weight updates) on a single sample. The speedup for the different configurations are shown with respect to this baseline. Since filter weight updates are only done once per batch, we show 2 speedup factors for a baseline batch size of 1, where weight communication overhead dominates, and for infinite batch size where weight update cost is negligible compared to other components and excluded. The actual speedup should be between these 2 depending on the batch size.

We observe that computation times dominate for small number of devices and tiles, but scale down with increasing number of devices and cores (more tiles). Due to the shared memory implementation within devices, there is no overhead for communication between tiles on the same device. Consequently, the communication overhead is uniform across different numbers of cores with the same number of devices. However, boundary data and weight communication overhead increases with more devices, where overhead for weight updates dominates for a larger number of devices. This limits speedup for small batch sizes and can outweigh savings in computation times, where 6 devices perform worse than 4. At the same time, we do observe strong scaling in the speedup for large batch sizes. We will further analyze results with varying batch size later.


\subsection{Memory}

\begin{figure}[tp]
    \centering
    \includegraphics[scale=.20]{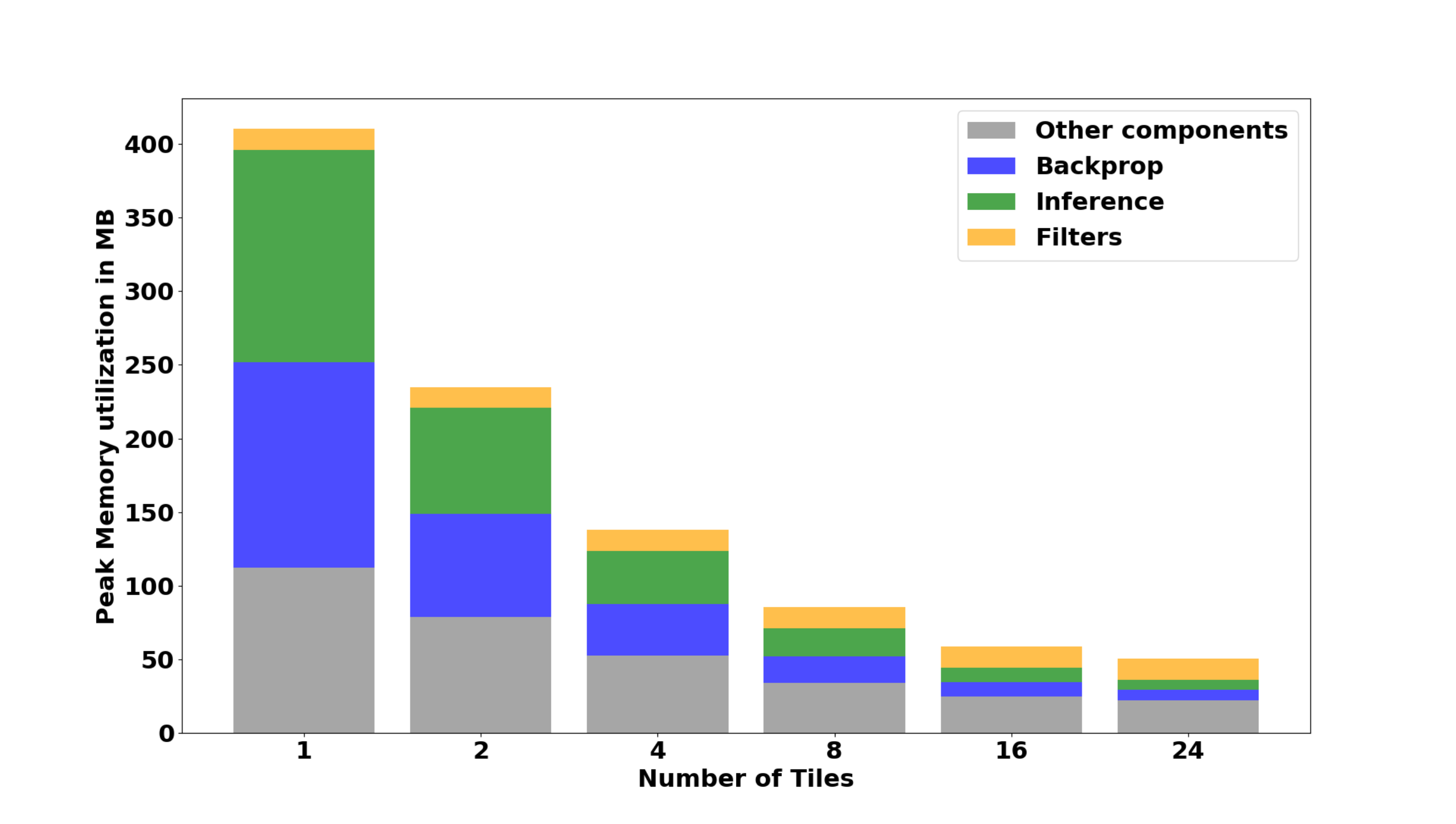}
    \vspace{-10pt}
    \caption{Memory utilization with number of tiles.}
    \label{fig:memory}
\end{figure}

Fig.~\ref{fig:memory} shows the peak physical memory utilization per tile measured while the training cycle of a single sample on the Raspberry-Pis was in progress. The figure also shows the split of the major memory usage components - feature maps, delta maps, filters and other implementation-related components such as a preallocated buffer for intermediate computation, communication buffers and code space.  The memory consumption is $\sim$400 MB per tile and drops to $\sim$50 MB per tile when using 24 tiles. In general, by tiling in a finer granularity, memory requirements per tile and hence per device are reduced. However, while memory requirements for feature, delta maps and other buffers decreases linearly with the number of tiles, filter memory usage is constant leading to diminishing returns.

\subsection{Batching and Grouping}

\begin{figure}[tp]
    \centering
    \includegraphics[width=0.95\textwidth]{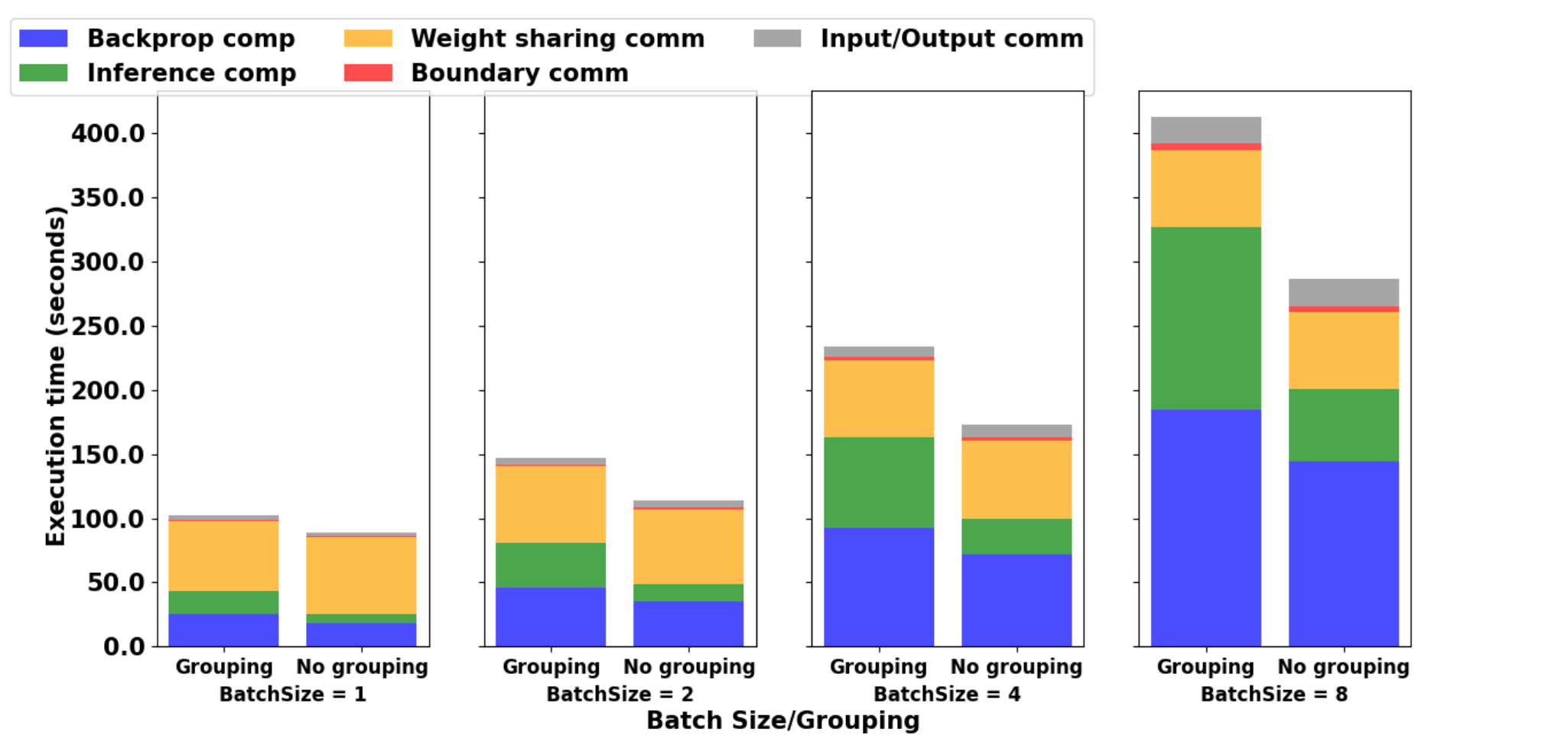}
    \vspace{-10pt}
    \caption{Comparison with batch size and grouping.}
    \label{fig:batch_grouping_comparision}
\end{figure}

We further conducted experiments on a batch of samples of various sizes. We also performed a comparison between grouping profiles - with and without grouping. Fig.~\ref{fig:batch_grouping_comparision} shows the result of running the training cycle on a batch size of 1 to 8 samples. We conducted this experiment using all 4 cores on the 6 Rapberry Pi devices using 24 tiles. We observe that synchronizing every layer (no grouping) performs significantly better than with grouping across all batch sizes. In case of the Raspberry Pis, total execution time is dominated by computation times or weight updates, and the improvement comes from optimization of computation. Computation costs scale proportionally with the number of samples in the batch, but the filter updates are done once per batch and take roughly the same time across batch sizes. As such, the relative contribution of weight update costs decreases with larger batches. At the same time, the boundary communication and input/output communication overhead increases with larger batch size, but is negligible compared to computation cost. Overall, Raspberry-Pi devices are computation limited and hence the synchronizing at every layer to minimize redundant computation is optimal.

\subsection{GPU experiments}
We also conducted experiments on a pair of Nvidia-Jetson Nano boards to illustrate the case of a communication-limited setup. Each board had a quad-core ARM Cortex-A57 CPU and a Maxwell architecture GPU with 128 CUDA cores. The 2 boards were connected using a 10Gbps Ethernet link. 

\begin{figure}[tp]
\centering
    \includegraphics[width=\textwidth]{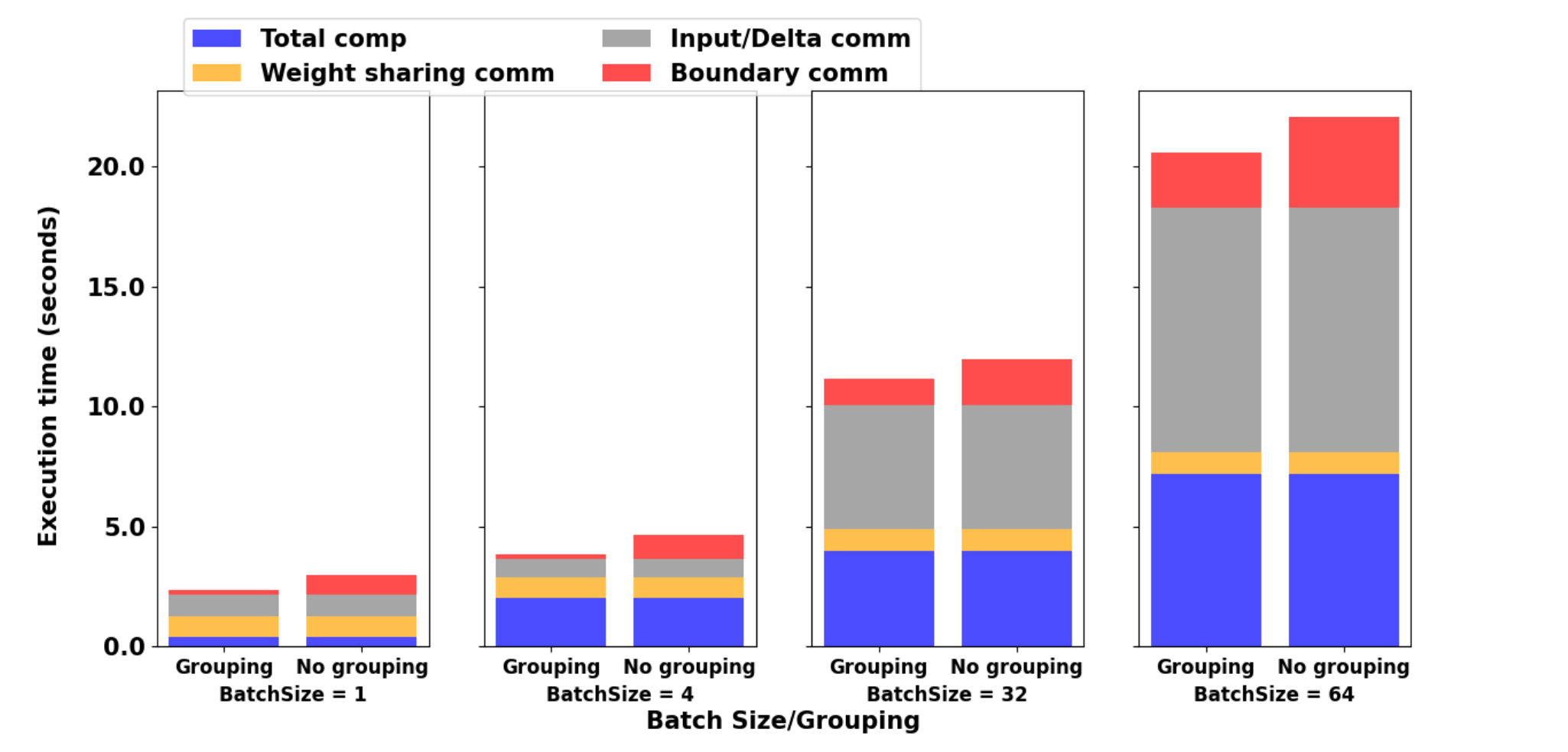}
    \vspace{-20pt}
    \caption{2-tile experiment with GPUs.} \label{fig:gpu_orig_comm}
\end{figure}

Fig.~\ref{fig:gpu_orig_comm} illustrates the single batch training cycle time for different batch sizes for a 2-tile setup (each board training on the GPU). On the GPUs, the inference plus backprop computation is much faster than on the Pis, thus making communication and synchronization overhead the limiting factor. In this case, the difference in boundary communication overhead among different groupings though small is noticeable. The case of with grouping performs better than without grouping since it synchronizes less frequently. On the GPUs, the extra redundant computation in the grouping case has negligible effect on computation time. By contrast, the extra communication and frequent synchronization, which includes transferring data to and from the GPU incurs a relatively larger overhead. Hence, it is more optimal to synchronize less frequently, and grouping is optimal. 

\section{Summary and Conclusions}
In this paper, we proposed a method for distributed mobile and edge training in feature-map dominated convolutional and pooling layers. Our method exploits locality in convolutional layers to partition feature maps and the delta gradients in forward and backward passes. It parallelizes training at a granular level within each sample. All intermediate layers are fused in that the core feature maps and delta maps remain local on the device with only a small overhead of shared data communication between neighboring tiles. Layers are further grouped based on a grouping profile that affects tradeoffs between computation, shared boundary communication and synchronization overhead. A grouping optimization algorithm including cost model and additional results are discussed in~\cite{techreport}. 
A reference implementation of our approach is available at~\cite{repo}. 
Future work will explore weight partitioning techniques and how to extend our approach to other weight-dominated layers.


%
%
%
%

\bibliographystyle{IEEEtran}
\bibliography{refs}

\end{document}